\documentstyle[prd,aps,floats,twocolumn]{revtex}

\def\P{{\cal P}}

\def\be{\begin{equation}}
\def\ee{\end{equation}}
\def\bea{\begin{eqnarray}}
\def\eea{\end{eqnarray}}

\renewcommand\({\left(}
\renewcommand\){\right)}
\renewcommand\[{\left[}
\renewcommand\]{\right]}

\newcommand\lsim{\mathrel{\rlap{\lower4pt\hbox{\hskip1pt$\sim$}}
    \raise1pt\hbox{$<$}}}
\newcommand\gsim{\mathrel{\rlap{\lower4pt\hbox{\hskip1pt$\sim$}}
    \raise1pt\hbox{$>$}}}

\def\P{{\cal P}}

\newcommand\sub[1]{_{\rm #1}}

\newcommand\pa{\partial}

\begin{document}
\preprint{} 
\draft

%
%
\input epsf
\renewcommand{\topfraction}{0.99}
\renewcommand{\bottomfraction}{0.99}
\twocolumn[\hsize\textwidth\columnwidth\hsize\csname@twocolumnfalse\endcsname

\title{Generating the Curvature Perturbation at the End of Inflation in String 
Theory}
\author{David H. Lyth$^1$ and Antonio Riotto$^2$}

\address{(1) Physics Department,
Lancaster University, Lancaster LA1 4YB,UK}
\address{(2) CERN, Theory Division, Geneve 23, CH-1211 Switzerland\\
INFN, Sezione di Padova, Via Marzolo 8, 35131, Italy}
\date{\today}
\maketitle
\begin{abstract}
\noindent
In  brane inflationary scenarios, the 
cosmological perturbations are  supposed to originate
from the vacuum fluctuations of the inflaton 
field corresponding 
to the position of the
brane. We show that a
 significant, and possibly dominant, contribution
to the curvature perturbation is generated at the end of inflation
through the vacuum fluctuations of fields, other than the inflaton, which
are light during the inflationary trajectory and become heavy at the 
brane-anti brane annihilation. These fields appear generically 
in string compactifications
where the background geometry has exact or approximate isometries and
parametrize   the internal angular directions of the brane.  
\end{abstract}


\vskip2pc]


\noindent
Significant progress has been recently 
made in obtaining brane inflation 
\cite{Dvali:1998pa,Burgess:2001fx,Dvali:2001fw}
from specific 
string theory constructions.
In these models, the distance between a $D$-brane and an anti $D$-brane
plays the role of the
inflaton. 

Flux compactification in type IIB string theory has become an ideal set up
to realize brane inflation. 
This is not only because it provides a promising
mechanism to stabilize the moduli fields 
\cite{Giddings:2001yu,Kachru:2003aw,Grana:2005jc} and therefore
a consistent string theory, but also because
the flux induced warped space has become a very interesting new ingredient
for model building. 

Warped throats provide also different mechanisms to achieve brane
inflation. Anti-D3-branes naturally settle down in throats. 
The warped geometry 
red-shifts the brane tension and reduces (easily by a large factor) the
mutual attraction between the branes and antibranes. In the 
KKLMMT scenario \cite{Kachru:2003sx,Firouzjahi:2005dh}, for instance, 
the entire physics of inflation takes place in the throat region and
it is mostly insensitive to the details of the ultra-violet Calabi-Yau
region. In the KKLMMT an anti $D$3-brane is sitting at the infra-red end 
 of the throat and a $D$3-brane is moving toward it. In the absence of 
anti branes, 
the $D$3 is a supersymmetric BPS object and it feels no force \cite{KS}. 
The only source of supersymmetry breaking in the compactification
is the inclusion of the anti $D$3-brane.
The insertion of a distant anti $D$3-brane just 
produces a very flat potential for the scalar field corresponding 
to the position of the
$D$3-brane, which plays the role of the inflaton field. Inflation ends 
when the branes come close together
and annihilate, allowing  the Universe to settle down to the string 
vacuum state that describes our Universe.

In the brane inflationary scenario, the 
curvature perturbation $\zeta$ is supposed to originate
from the vacuum fluctuations of the inflaton 
field itself. In this short note, we point out a new possibility; that
a significant, and possibly dominant, contribution
to the curvature perturbation is generated at the end of inflation
through the vacuum fluctuations of fields, other than the inflaton, which
are light during the inflationary trajectory and become heavy at the 
brane-anti brane annihilation. Indeed, 
if the branes are embedded in a string compactification
where the background geometry has exact or approximate isometries, there are
scalar brane degrees of freedom  transverse to the mutual distance and
parametrizing  the internal angular directions of the mobile brane. Their
potential is so flat that they have practically no influence on the inflaton
trajectory. However, being  
almost massless during the de Sitter epoch,  
their   quantum fluctuations are
excited with a nearly scale-invariant spectrum.  

Inflation ends when the mobile brane and the anti brane
 come close together
and annihilate. More specifically, 
at each position in space, inflation ends when the inflaton
field $\phi$ has some critical value $\phi_{\rm e}$. Usually, it is
assumed that the end of inflation is controlled entirely by the inflaton, so 
that $\phi_{\rm e}$ is independent of the position in space. Then inflation
ends on a spacetime slice of uniform energy density. However, 
the value $\phi_{\rm e}$ depends on the internal angular directions of 
the branes and, if these these phases are quantum
mechanically excited during
inflation, the value $\phi_{\rm e}$ will depend upon the position in space.

Under
these circumstances, there is no reason why inflation
should end on a slice of uniform energy density and one expects
the  comoving curvature perturbation $\zeta_{\rm e}$ to be generated 
at the end of inflation
 \cite{bu,bku,lythend,al} when  the light fields become very massive
and decay (after the end of the inflation the branes have 
disappeared and certainly cannot fluctuate anymore).

As a specific and concrete 
example, let us discuss again the KKLMMT model for $D$-brane 
inflation. 
The KKLMMT scenario corresponds to a Calabi-Yau geometry that develops
a long warped AdS-like throat and it corresponds to a
compactification of the Klebanov-Strassler (KS) solution \cite{KS}.
The KS solution  has an 
$SU(2)\otimes SU(2)$ isometry group. 
For example, the ultra-violet geometry 
of the  throat is that of a cone over the Einstein manifold 
$T^{1,1}=SU(2)\otimes SU(2)/ U(1)$. 
In the infra-red region the singular tip
of the cone is smoothen out to a three-sphere, which still preserves
an $SU(2)\otimes SU(2)$ symmetry. 

During its motion toward the
infrared, the $D$3-brane can fluctuate in the internal angular directions.
Since during the inflationary period, the two brane set up 
approximately preserves the $SU(2)\otimes SU(2)$ symmetry, 
the scalar fields associated with the angular positions are almost
massless and lighter than the Hubble rate. Furthermore, being angles, 
they do not get masses from the volume stabilization
mechanism. In the K\"ahler potential
of four-dimensional supergravity the volume modulus $\rho$ always appears in 
the
combination $\rho+\bar\rho - K(\phi_\alpha,\bar\phi_\alpha)$ \cite{GD}, 
where $\phi_\alpha$ collectively
denote the position of the branes in the six internal directions,
and $K$ is the K\"ahler potential for the geometry.
This coupling generates a mass for the fluctuations $\phi_\alpha$.
However, since the geometry has isometries, 
$K(\phi_\alpha,\bar\phi_\alpha)$ 
does not depend on some of the angles. In this way, the 
form of potential for the angular
fluctuations is not affected by the stabilization mechanism and leaves
some angles much lighter than the Hubble rate during inflation.
At the end of inflation, the $D$3-brane comes close to the anti $D$3-brane.
At stringy distances, we cannot ignore the fields coming from
open strings connecting the branes. In particular, in a generic 
brane-antibrane system, the open string ground state is a tachyon $T$ 
with mass of order $M_s$,
the string scale.

Since the annihilation takes
place at the bottom of the throat, the tachyonic mass is redshifted to
$m_T^2\sim -a_0^2M_s^2$, where 
 $a_0$ is the redshift
factor at the bottom of the throat. Unfortunately, 
the precise potential is rather dependent on the 
detailed structures of the compactification and remains 
to be explored more carefully. However, to illustrate our point, 
we can  write 
a  potential for the inflaton $\phi$, the tachyon $T$ and the
angular positions $\theta_\alpha$  which captures the salient 
features during inflation and 
of the tachyonic condensation when the brane and anti brane
are close to each other. Keeping for illustrative purposes 
just one angular position,  we write
\begin{eqnarray}
V(\phi,T, \theta)&=&\frac{m^2}{2}\phi^2 + 
T_3 a_0^4 \[ 1-c \( \frac{a_0 M_s}{\phi} \)^4 \]
\nonumber\\
&+&
T^2\left[\phi^2-a_0^2M_s^2(a + b \cos\theta ) \right]+\cdots\,  ,
\label{fullpot}
\end{eqnarray}
where $T_3$ is the tension of a $D$3 brane and 
the ellipses account for 
higher-order power terms in the field $T$ and of
possible terms responsible for tiny masses for the phase $\theta$.  
One  expects $a\sim b\sim 1$,  and the coefficient $c= (1/MK)\ll 1$, 
 where $M$ and $K$ are large
integers parametrizing the flux units.

The second line  vanishes during inflation, and specifies the inflaton
field value $\phi_{\rm e}$ at the end:
\be
\phi_{\rm e}^2  = a_0^2M_s^2(a + b \cos\theta )\, .
\label{phie}
\ee
The quadratic term for the inflaton field $\phi$ 
 receives contributions from a number of sources and is rather 
model-dependent \cite{Kachru:2003sx,sh}. 
However $m^{2}$ is expected to be comparable 
to $H^2=V_{0}/3M_{p}^{2}$, where $M_{p}$ 
is the reduced Planck mass ($G^{-1}=8 \pi M_{p}^{2}$) (this might 
represent an   obstacle in getting a sufficient number of e-foldings, but 
various ways to solve this problem
have been proposed \cite{sol}).

As long as
$\phi$ is much larger that the (redshifted) string scale, 
$T$ has a large positive 
mass, its vacuum expectation value $\langle T\rangle$ vanishes
 and all the terms in the potential involving $T$ can be ignored.
The inflationary potential is then independent of the angle $\theta$. 
At stringy distances, $T$ becomes tachyonic
and triggers the brane-anti brane annihilation. When $T$ condenses,
all the fluctuation fields on the
branes acquire masses roughly of order of the redshifted string scale. 
These masses are ${\cal O}(a_0M_s)$ and much  larger than the value of
the Hubble rate $H\sim a_0^2M_s$ at annihilation. 
All these massive fields will eventually decay. 

There are two effects that break the $SU(2)\otimes SU(2)$ invariance and 
may prevent the angular fluctuations of the brane 
from being exactly massless during inflation \cite{zaffa}. 

There is a small source of breaking
of the  $SU(2)\otimes SU(2)$ symmetry due to the presence of the anti
$D$3-brane
in the infra-red, which sits at a specific point on the three-sphere.
This explicit breaking is though suppressed when the 
the distance between the two branes during the inflationary
period is sizable.
The second
effect is caused by the moduli stabilization that
requires fluxes in the ultra-violet region typically breaking
$SU(2)\otimes SU(2)$; it is well known indeed that a Calabi-Yau manifold
has no isometries at all. Even this second effect is suppressed by the
distance. An estimate 
shows that the typical mass induced by the $SU(2)\otimes SU(2)$ breaking
scales as $a^8(\phi)$ \cite{giant}, where $a(\phi)$ is 
the warp factor evaluated at the position $\phi$ of the brane. This mass 
contribution is therefore  
 suppressed compared to the Hubble rate during inflation.

There is 
a  dual interpretation of these
massless fields. As familiar from the AdS/CFT correspondence and the
holographic interpretation, the throat part of the 
compactification  can be effectively replaced by a strongly interacting
gauge theory coupled to the four-dimensional gravity. In this picture,
the $D$3 brane position corresponds to a flat direction in the moduli space
of vacua of the gauge theory.
The global symmetry $SU(2)\otimes SU(2)$ is spontaneously broken along
this flat direction \cite{comment}.
The angular
positions of the brane are therefore 
identified  with the massless Goldstone bosons of 
the spontaneously  broken global symmetry. 

Now we turn to the calculation of the curvature perturbation. 
The phase will be
 $\theta\simeq \sigma/\phi_{\rm e}$,
where $\sigma$ is canonically normalized. Using the $\delta N$-formalism, we 
may compute the  spectrum of the curvature
perturbation  \cite{lythend,al}
\bea
\label{a}
\P_\zeta&=& \[ \(\frac{\pa N}{\pa \phi} \)^2 + 
\(\frac{\pa N}{\pa \sigma} \)^2 \] \( \frac{H}{2\pi} \)^2\, ,  \\
\label{aa}
 \(\frac{\pa N}{\pa \phi} \)^2 &=& \frac1{M_p^2} \left.
\frac {V}{V'} \right|_* \, ,\\
\label{aaa}
\(\frac{\pa N}{\pa \sigma} \)^2 &=& \frac1{M_p^2} \left.
\frac {V}{V'} \right|_{\rm e}  \( \frac{d\phi_{\rm e}}{d\sigma} \)^2\, .
\eea
In Eq. (\ref{a}), the first piece corresponds to the curvature perturbation
generated during inflation, while the second
to the curvature perturbation $\zeta_{\rm e}$ 
generated at the end of inflation.
It coincides with the perturbation of the change of the number of
e-folds from a spacetime slice of uniform energy density just before the end
of inflation to a spacetime slice of uniform energy density just after
the end of inflation. 
In Eqs. (\ref{aa}) and (\ref{aaa}),  $V(\phi)$ 
is the potential during inflation and  $3M_p^2H^2 =V
\simeq T_3 a_0^4$. The  
star denotes the epoch when cosmological scales leave the horizon,
and the  subscript e denotes the end of inflation.
For a typical value of $\theta$,  we have
$\phi_{\rm e} \sim a_0 M_s$ and $d\phi_{\rm e}/d\sigma\sim 1$.

The first contribution is generated at horizon exit
by the fluctuation of the inflaton,
and   the second contribution is generated at the 
end of inflation by the fluctuation of the  phase field.
(We assume  that $\zeta$ remains constant afterward.)
Let us see if the latter contribution can dominate.
This will not happen if 
  the mass term is negligible,  because then 
  $V'/V\propto \phi^{-5}$ is increasing with time. 

Suppose  instead that the mass term dominates throughout inflation.
This  is the case if 

\begin{equation}
c\ll \eta \phi_e^2/M_p^2\, ,
\end{equation}
where $\eta\equiv m^2/3H_*^2$.
Then  we have $V'/V  \propto \phi$ and

\begin{equation} 
\phi_* = e^{N\eta} \phi_{\rm e}\, .
\end{equation}
We are now dealing with  the model investigated in \cite{al},
using a somewhat different notation.
For typical values of $\theta$, the 
 fluctuation of the phase field
dominates if $\eta N\gsim 1$. Then the  spectrum 
has the observed value if
\be
5\times 10^{-4} \sim \frac{\sqrt{T_3} a_0}{\eta M_s M_p}\, ,
\ee
which can be achieved by a suitable choice of parameters.

As $c$ is increased with the other parameters fixed,
 there is a continuous transition to the unviable 
case that the mass term is negligible throughout inflation.
This transition is complete when
$c \sim  \eta \phi_{\rm e}^2 \sqrt{\eta N}/M_p^2$.

The spectral index of the curvature perturbation
is determined by  the potential of $\sigma$
evaluated at horizon exit
through the  expression  \cite{ss,treview}
$n=1+ 2V''(\sigma)/3H^2$. The potential is roughly 
$m_\sigma^2 \phi_*^2 \cos(\sigma/\phi_*)$, and if
 $m_\sigma^2/H^2$ is tiny as we suggested earlier,
$n$ is indistinguishable from 1, which is  consistent with 
 observation at the 2- to 3-$\sigma$ level \cite{sigmas}.

Finally, we consider non-Gaussianity with the mass term dominating.
Assuming that the fluctuation of the phase field dominates,
the  parameter for the primordial non-Gaussianity is
\be
\frac35f\sub{NL} = 
\frac{\pa^2 N}{\pa \sigma^2} \( \frac{\pa N}{\pa \sigma}\)^{-2}
\sim \eta \frac{\phi_{\rm e}}{M_p} 
 \frac{d^2 \phi_{\rm e}}{d \sigma^2}   
\( \frac{d\phi_{\rm e} }{d \sigma} \)^{-2}\, .
\ee
We  need $|f\sub{NL}|\gsim  1$ for it to be eventually observable 
\cite{komatsu}. As
was investigated in \cite{al}, that may be possible for a very special
choice of the phase angle, but there 
seems to be no reason to make such a choice.
In particular, anthropic selection does not seem to 
 favour (or disfavor) significant non-Gaussianity \cite{lythcurv}.

To summarize,  we have pointed out that in inflationary scenarios
built up in string models, where inflation is driven by mobile branes moving
in certain compactification manifolds, the cosmological perturbations
may be 
generated at the final stage of inflation, when brane and anti brane 
annihilate, through the vacuum fluctuations of pseudo Nambu-Goldstone fields.
They  are associated to the  isometries  present when the brane and anti brane
are far apart and disappear close to the tachyonic point. The
generation of the curvature perturbation in such a case is due to the
fact that the end of inflation happens in different space points
at different instants of time.
As a specific example we have discussed the 
KKLMMT scenario, but we expect  our findings to 
be quite generic.



\begin{references}
\bibitem{comment} The set up with 
K and M fluxes is dual to a $SU(KM+M)\otimes SU(KM)$ $N=1$ supersymmetric
theory with pairs of bifundamental fields $A_i$ and $B_i$, $i=1,2$,
each one rotated by a global $SU(2)$ symmetry, with an $SU(2)\otimes SU(2)$
invariant superpotential. The same compactification with a $D$3 brane
inserted can be represented by a gauge group
 $SU(KM+M+1)\otimes SU(KM+1)$ broken to $SU(KM+M)\otimes SU(KM)$
along a flat direction where $A,B$ acquire a VEV breaking the 
global symmetry $SU(2)\otimes SU(2)$.
\end{references}
\end{document}